\documentclass[sigconf,nonacm]{acmart}

\usepackage{gensymb}
\usepackage{array}
\usepackage{dblfloatfix}    
\usepackage[absolute,overlay]{textpos}

\AtBeginDocument{%
  \providecommand\BibTeX{{%
    \normalfont B\kern-0.5em{\scshape i\kern-0.25em b}\kern-0.8em\TeX}}}

\begin{document}

\thispagestyle{empty}

\title{Open5x: Accessible 5-axis 3D printing and conformal slicing}


\author{Freddie Hong}
\email{t.hong19@imperial.ac.uk}
\orcid{}
\affiliation{%
  \institution{Imperial College London}
  \streetaddress{}
  \city{London}
  \state{}
  \country{United Kingdom}
  \postcode{SW7 2AZ}
}

\author{Steve Hodges}
\email{steve.hodges@microsoft.com}
\orcid{}
\affiliation{%
  \institution{Microsoft Research}
  \streetaddress{}
  \city{Cambridge}
  \state{}
  \country{United Kingdom}
  \postcode{}
}

\author{Connor Myant}
\email{connor.myant@imperial.ac.uk}
\orcid{}
\affiliation{%
  \institution{Imperial College London}
  \streetaddress{}
  \city{London}
  \state{}
  \country{United Kingdom}
  \postcode{SW7 2AZ}
}

\author{David Boyle}
\email{david.boyle@imperial.ac.uk}
\orcid{}
\affiliation{%
  \institution{Imperial College London}
  \streetaddress{}
  \city{London}
  \state{}
  \country{United Kingdom}
  \postcode{SW7 2AZ}
}

\renewcommand{\shortauthors}{    }

\begin{abstract}
  
  
  
 The common layer-by-layer deposition of regular, 3-axis 3D printing simplifies both the fabrication process and the 3D printer's mechanical design. However, the resulting 3D printed objects have some unfavourable characteristics including visible layers, uneven structural strength and support material. To overcome these, researchers have employed robotic arms and multi-axis CNCs to deposit materials in conformal layers. Conformal deposition improves the quality of the 3D printed parts through support-less printing and curved layer deposition. However, such multi-axis 3D printing is inaccessible to many individuals due to high costs and technical complexities. Furthermore, the limited GUI support for conformal slicers creates an additional barrier for users. To open multi-axis 3D printing up to more makers and researchers, we present a cheap and accessible way to upgrade a regular 3D printer to 5 axes. We have also developed a GUI-based conformal slicer, integrated within a popular CAD package. Together, these deliver an accessible workflow for designing, simulating and creating conformally-printed 3D models.
 
\end{abstract}





\maketitle

\section{Introduction}
Additive manufacturing (AM) techniques, and in particular 3D printing, have become very popular for fabricating physical prototypes in various industries owing to the ease of constructing bespoke and complex geometries on demand. In 3D printing, objects are typically constructed in a layer-by-layer deposition. Among various approaches, fused filament fabrication (FFF, also known by some as fused deposition modelling or FDM) is the most popular form of 3D printing for individual makers due to its low barrier of entry: machines are low cost, small and easy to use. 

Within HCI research, various techniques and tools have been introduced to expand the capability of desktop FFF 3D printing. Researchers have implemented tool changing mechanisms to combine various tools and materials within a single build environment \cite{10.1145/3313831.3376425,Teibrich2015PatchingObjects}, developed new methods of multi-material 3D printing \cite{Takahashi2020ProgrammableFilament} and developed custom software for various novel approaches to 3D printer toolpath generation \cite{Mueller2014WirePrint:Previews}.

While the planar layer construction of FFF simplifies the mechanical design of the 3D printer and the computation of the toolpath, the finished products often exhibit unfavourable characteristics such as visible layers, rough surface finishes, uneven mechanical properties and the inclusion of sacrificial support material. To expand the capabilities of 3D printing, various methods of depositing materials in conformal layers have been explored. For example, researchers have employed 6 degree of freedom robotic arms \cite{Bhatt2020,Dai2018Support-freeMotion,Fry2020RoboticOrientations} and multi-axis computer-numerical control (CNC) \cite{Gardner2018} to deposit the materials in curved layers. In particular, 5-axis 3D printing offers a variety of advantages over conventional 3-axis method, including support-less printing \cite{Bhatt2020}, curved layer deposition that strengthens the mechanical properties of the 3D printed object against multi-directional loads \cite{Fang2020ReinforcedStrength, Gardner2018, Luo20203DParts}, conformal surface finishes \cite{Shembekar2018} and direct printing of functional materials on conformal surfaces \cite{Urasinska-Wojcik2019IntegratedApplications}. However, 5-axis 3D printing is currently out of reach for many users
because: i) software that allows users to prepare the 3D files for 5-axis printing is not readily available, ii) many individual makers cannot afford the necessary hardware, and iii) the machines take up significant amounts of space and are often not suitable for personal fabrication.

In this work we take inspiration from and briefly review the latest AM research and various creative design techniques showcased in HCI research for personal fabrication. We show how a popular off-the-shelf 3-axis 3D printer can be upgraded to 5-axis 3D printing, with the aim of empowering more HCI researchers and makers to take advantage of the benefits of conformal 3D printing. An online repository contains details of all the mechanical parts and 3D printing files needed to perform this upgrade.
We present our GUI-based design software that allows users to generate the necessary conformal toolpaths
and we present several scenarios that showcase the benefits of 5-axis 3D printing.
\section{Related work}

\subsection{Conformal 3D printer}
There have been several approaches to using multi-axis CNC machines and 6 degree of freedom robotic arms as FFF printers. For instance, Bhatt et al. \cite{Bhatt2020b} and Dai et al. \cite{Dai2018Support-freeMotion} employed filament extruder and a dynamic built platform mounted at the tip of the robotic arm to construct support-free 3D printed object. Similarly, Gardner et al. \cite{Gardner2018} and  Kaill et al. \cite{Kaill2019ALoading} used 5-axis CNC with filament extruder to extrude filament in conformal layers. While these multi-axis machines are capable of conformal 3D printing, they are not originally designed for 3D printing and thus the size and cost of the machine exceed the capacity of the most individual makers. Researchers have alternatively constructed more `desktop friendly' 5-axis 3D printers. For instance, Isa et at. \cite{Isa2019Five-axisLayers} and Zhao et al. \cite{Zhao2020ACLFDM} designed and constructed a new custom 3D printer that is dedicated to conformal 3D printing. In a different approach, Teibrich et al. \cite{Teibrich2015PatchingObjects} developed a hardware setup that is integrated into an existing desktop 3D printer to combine subtractive and additive manufacturing on a rotary platform. Their system, however, does not allow for synchronised movement in 5 axes, and is therefore not capable of conformal 3D printing.

To make 5-axis 3D printing more easily accessible to a broader community of makers, we present a method of converting an existing cartesian 3D printer into a 5-axis 3D printer by replacing the print bed with controllable rotating stage. We include a supplementary repository wherein we openly provide our 3D design CAD files, electronics schematics, 5-axis firmware and step-by-step guide to install the 5-axis gantry to a cartesian 3D printer. We have developed our work using a Prusa i3 MK3s, but the design can be modified to suit other brands of desktop 3D printers.

\subsection{Conformal slicing}
One of the biggest challenges involved in multi-axis 3D printing is the G-code generation, which requires complex computation of the toolpath and machine control compared to conventional 3D printers. Various researchers have developed their own algorithms for controlling the 5-axis printer \cite{Shembekar2018,Feng2021Curved-layeredMachine}, but there is limited development in terms of user-friendly conformal slicers that general users without the scripting knowledge can employ. This contrasts to the abundance of planar slicers for planar 3D printers that can easily generate 3D printing G-code with desirable settings, such as the print speed, layer height and infill density. Similarly, using 5-axis CNC for milling and routing has also been around for a long time and there are many existing works for optimizing the toolpath and machine control \cite{Yang2015ATheory,Sencer2008FeedConstraints}. As such, we took inspiration from the works of 5-axis milling computation for developing our own conformal slicers and firmware dedicated for conformal 3D printing.  

While there are researchers who compute the conformal slicing and toolpath generation through numerical computing programs such as MATLAB (MathWorks, Natick, Mass), there are also groups of researchers who use 3D modelling software with visual scripting extensions, namely Rhinoceros and Grasshopper \cite{Grasshopper}. For instance, Lim et al. \cite{Lim2016ModellingComponents} and Gardner et al. \cite{Gardner2018} used Grasshopper to visualise the conformal toolpath and produce the G-code. Visual scripting integrated into a CAD software allow the user to design the 3D model and slice the model together in a single digital environment. This process is also more accessible to the general user without previous knowledge in numerical scripting. In this paper, we developed our own conformal slicer based on the visual scripting language with added graphical user interface (GUI) elements such that it is as close as possible to plug and play for users with no background in scripting.

\subsection{Personal fabrication}
In personal fabrication, machine design takes into account that the end users may be individual makers who may have limited access to space and resources. 
Various approaches to expand the capabilities of desktop FFF 3D printers have been proposed with this in mind. 
For instance, \textit{Jubilee} \cite{10.1145/3313831.3376425} introduced an open source desktop tool changing machine that can combine 3D printing with other processes, such as plotting and syringe writing. Jubilee can be built with readily available parts and comprehensive assembly instructions are provided. \textit{WirePrint} \cite{Mueller2014WirePrint:Previews} introduced a method of 3D printing wireframe structures not in a layer-by-layer manner but through 3D printing in mid-air. This allows much faster printing compared to traditional layer-based printing, also giving the user a higher degree of freedom in controlling the extrusion direction to control the properties of the printed object. The proposed technique can be used on any kind of FFF based desktop 3D printers. Similar to above, we provide comprehensive instructions for both hardware and software to ease the on-boarding process to 5-axis 3D printing.

\section{5-axis System}
\subsection{Upgrading from 3-axis to 5-axis}
\begin{figure}[h]
  \centering
  \includegraphics[width=1.0\linewidth]{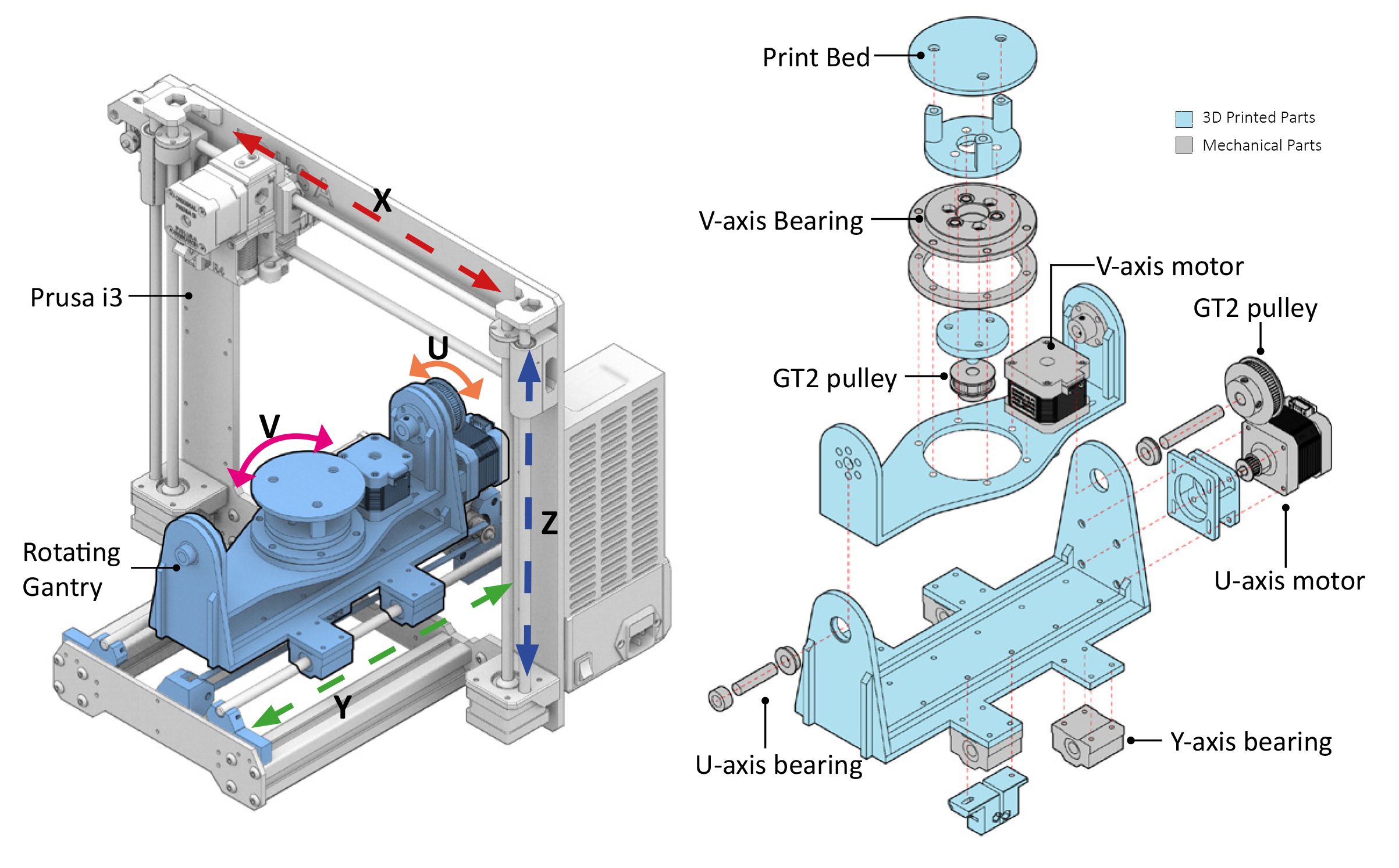}
  \caption{Left: Overall arrangement of the 5-axis 3D printer with the 2-axis mechanism built-into the existing Prusa i3 MK3s. The 2-axis rotary gantry is indicated in blue and the 5 axes (X,Y,Z,U \& V) are illustrated with colored arrows. Right: Exploded isometric drawing of the 2-axis rotary gantry. The 3D printed parts are cyan and the commercial parts are grey.}
  \label{arrangement}
  \Description{}
\end{figure}

The primary objective of this work is lowering the entry barrier of 5-axis 3D printing. Since desktop FFF 3D printers are widely available and common to designers and makers, instead of making a whole new machine we decided to augment an existing open-source desktop 3D printer, the Prusa i3 MK3s, into a 5-axis machine by adding a 3D printed 2-axis rotating gantry. This way the user can work with an already familiar hardware system while exploring 5-axis 3D printing. The overall hardware arrangement of the 5-axis upgrade is shown in left side of Figure \ref{arrangement}. The right side of Figure \ref{arrangement} shows the 2-axis gantry that is built with a mixture of 3D printed parts and readily-available mechanical parts, such as belt and pulleys.
A full list of the necessary parts can be found in our open repository\footnote{A link to the open repository will be included in the paper upon the acceptance of the work for publication.} along with CAD files for 3D printed parts. The position of the Y-axis linear bearing can be modified to suit different models and brands of 3D printers. In addition, we designed a replaceable print bed with various diameters. It is useful to match the print bed size as closely as possible to the base diameter of the intended 3D printed object to reduce the collision angle between the bed and the nozzle.

\subsection{Firmware and electronics}
\begin{figure}[h]
  \centering
  \includegraphics[width=1.0\linewidth]{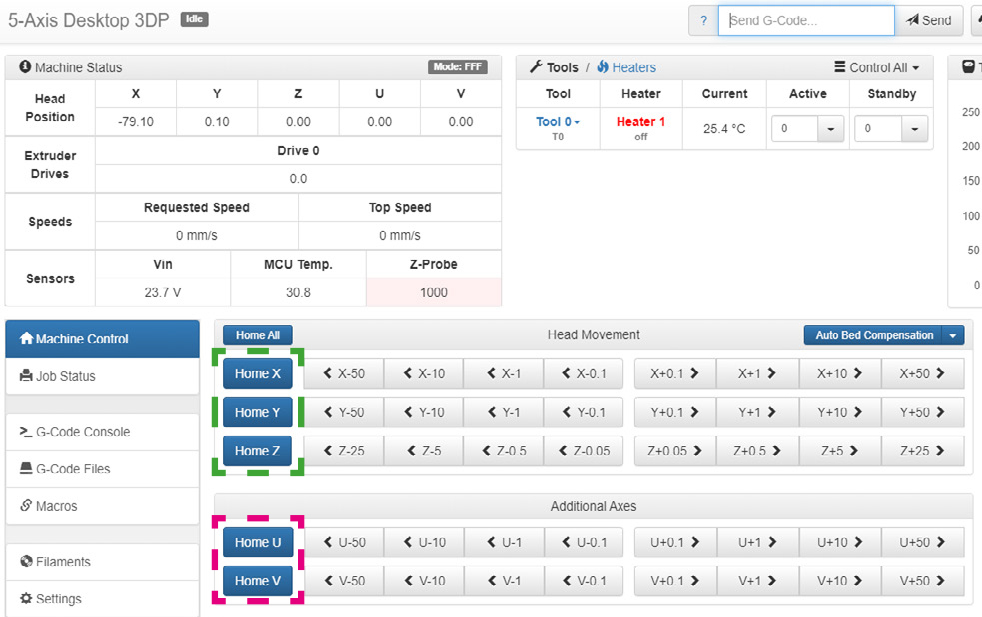}
  \caption{Duet web controller for 5-axis 3D printer. Cartesian axis are indicated by the green and the two rotary axis U and V are indicated by the pink}
  \label{dwc}
  \Description{}
\end{figure}

For our 5-axis conversion, we replaced the original Prusa i3 electronics board with the RepRap firmware compatible 3D printing board called Duet 2 \cite{Duet3DDuetOverview}. Duet 2 supports the addition of an expansion board that allows up to 5 additional stepper motors. The process of replacing the electronics board is straightforward: a user simply re-wires the existing electro-mechanical elements from the original Prusa i3 board to corresponding pins on the new Duet 2 board. We share full details on pin connections in our open repository.
To control Duet 2, we used RepRap firmware version 3.1.1. We implemented the original Prusa i3 machine profile with the RepRap firmware, and we added new machine profile and configurations to suit our 5-axis operation. Once the Duet 2 board is fitted, the user can simply upload our configuration file and start to explore the machine via the Duet web control \cite{DuetDuetControl} as shown in Figure \ref{dwc}.

\section{Conformal slicer}
Here we demonstrate our conformal slicing tool built on visual scripting language Grasshopper, which runs within the commonly used CAD software Rhinoceros 3D \cite{Grasshopper}, commonly referred to as Rhino. The benefit of using Grasshopper is that the user can construct the 3D file, produce the conformal toolpath, simulate the printing process and export the G-code all within a single digital environment. There are several parts to slicing: 1) generating the conformal toolpath, 2) computing 5-axis inverse kinematics and movement control, and 3) optimization of the 5-axis movement for 3D printing and exporting the machine commands in G-code ready to print.

\subsection{Toolpath creation}
\begin{figure}[h]
  \centering
  \includegraphics[width=1.0\linewidth]{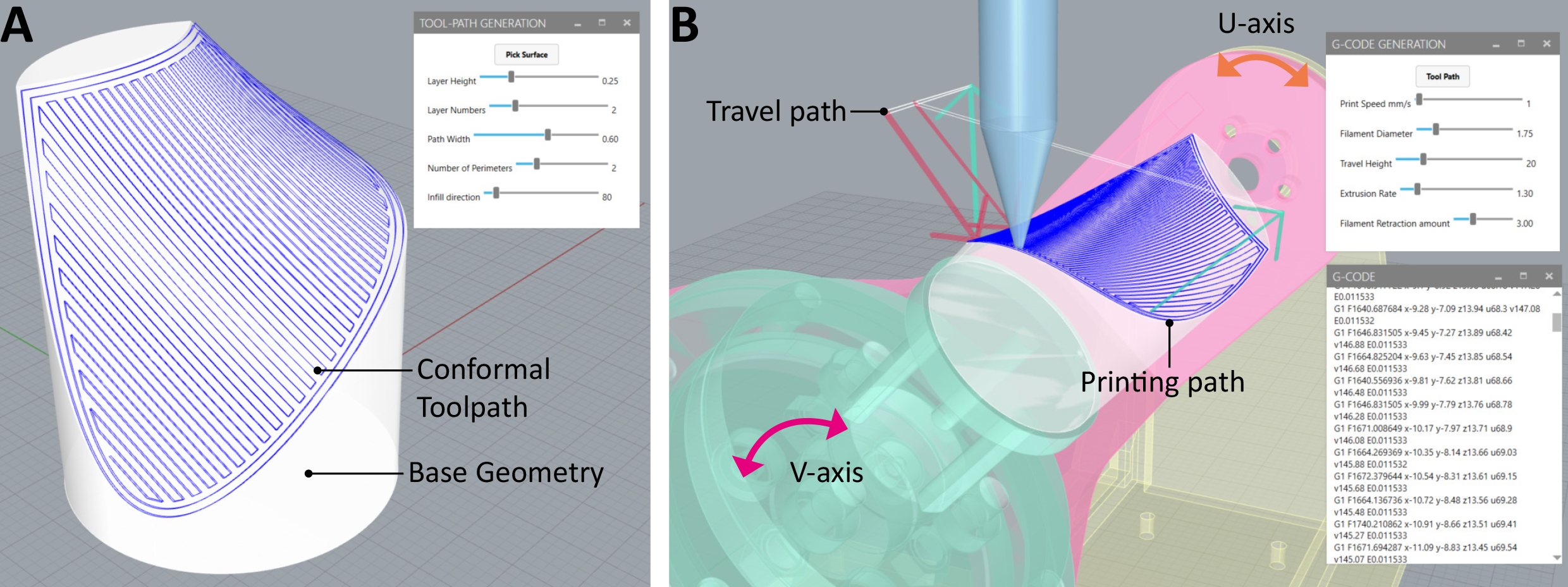}
  \caption{Parametric slicing tool integrated in Rhino3D: (a) conformal toolpath generator, and (b) 5-axis G-code generator}
  \label{slicing}
  \Description{}
\end{figure}

The process begins by importing the geometry into Grasshopper. Once imported, the user can either select the surface or a geometry that will be conformally 3D printed, and our editor automatically generates the toolpath onto the surface. Our slicing parameters include nozzle size, layer height, travel height, infill pattern direction and the number of perimeters. The infill and perimeter toolpath are drawn conformally to the substrate surface with B-splines. B-splines are then subdivided into polylines, which consist of segments of the straight lines. Unlike planar 3D printing, for 5-axis the nozzle is moving in 3D space during extrusion. Therefore, the length of each segment defines the resolution of the conformal print, in addition to the layer height. We set the segment length to be 0.2~mm, around half the size of a common 3D printing nozzle diameter. Users can change the slicing parameters and check the generated toolpath on-screen in Rhino. As shown in Figure \ref{slicing}, the GUI of our conformal slicer was intended to resemble the interface of common planar 3D printing slicers. To transition from one toolpath to the next, a retraction function is required. Our toolpath editor calculates the surface normal at the points where the paths start and end to determine the angle of retraction. The height of the travel can be manually adjusted if necessary to avoid any collision with the substrate.

\subsection{G-code generation}
\begin{figure}[h]
  \centering
  \includegraphics[width=1.0\linewidth]{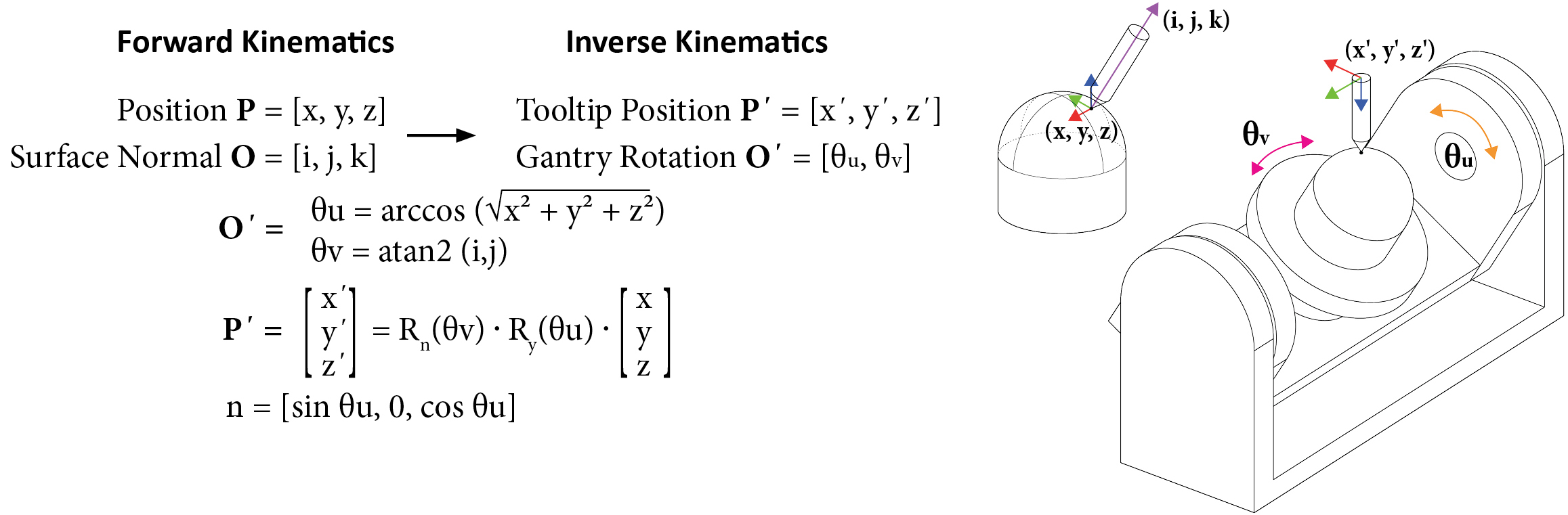}
  \caption{Translation of forward kinematics to inverse kinematics for 5 axis printing.}
  \label{matrix}
  \Description{}
\end{figure}

Once the conformal toolpath is created, the user can adjust the machine parameters such as print speed, non-print travel speed, extrusion rate and the amount of filament retraction. Calculating the machine commands---called G-codes---from the toolpath is more complex than for planar 3D printing due to the continuously changing Z axis. As described above, the toolpath is subdivided into segments of straight lines and points which defines the position of the nozzle tip. We first find the forward kinematics of the machine defined by position of the point on the surface described in Cartesian coordinates (X,Y,Z) and the unit surface normal at point described as (I,J,K), which is also based on Cartesian coordinates. Next, we use polar coordinates and rotational matrix to achieve our inverse kinematics, i.e., the rotation of our bed and position of the nozzle tip. As shown in Figure \ref{matrix}, we use \textit{arccos} and \textit{atan2} functions to find the rotational value of $\theta_u$ and $\theta_v$. We first rotate the original point by $\theta_u$ at y-axis and rotate again with $\theta_v$ at continuously changing V-axis. The direction of continuously changing V-axis can be found by [sin$\theta_u$, 0, cos$\theta_u$] described in vector coordinates. This two rotation returns the position of the nozzle tip P'.

\subsection{5-axis movement optimization}
Unlike planar 3D printing, the speed at which the nozzle travels on the surface of the substrate does not correlate to the speed of movement for each axis. For a movement between the two points in 5-axis the distance that all the axes need to travel could be much greater or smaller than the actual length of the segment. This means that when 3D printing in 5 axes, if all axes are set to operate at constant speed the resulting speed of the nozzle on the surface of the substrate will be continuously changing. This in turn impacts filament extrusion and has a negative impact on the quality of the printed object. In order to overcome this, for every segment over which the nozzle will travel, we calculate the ratio between the total distance the machine needs to travel and the length of that segment. We adopted speed optimisation calculations from the literature on 5-axis CNC milling \cite{Inc.2005FiveControls}. For our calculation of the optimal speed, we considered the rotary movements as linear motions and included them in the necessary Euclidean distance calculation. We first find the total Euclidean relative distance between segment for all axes including the extrusion using  $ d = \sqrt{\Delta x^2 + \Delta y^2 + \Delta z^2 + \Delta u^2 + \Delta v^2 + \Delta e^2}\ $, and then we divide this by the length of the segment $l$ to find the ratio between the travel distance and segment length. The optimized printing speed can be achieved by multiplying the desired printing speed with the speed compensation ratio, which is $F' = F \cdot \frac{d}{l}$. Our Grasshopper G-code editor also outputs the information on the speed of each axis at any point during the 3D print to ensure that speed of the axis does not exceed the mechanical threshold of the motor.

In addition to speed optimization, we also implement optimized rotation for the the print bed (V-axis). 

Bed rotation is specified as an angle between 0 to 360$^{\circ}$, but that sometimes prevents the machine from taking the shortest path; for instance, if to rotate from 355$^{\circ}$ to 1$^{\circ}$, by default a path of -354$^{\circ}$ will be generated instead of +6$^{\circ}$. This increases bed travel time and potential risks of collisions. To resolve this, we implemented a relative polar coordinate solution by calculating deltas between the absolute angular values. Then we process the new relative values through \textit{arcsin} and \textit{sin} functions to find the shortest paths.


For users, all of these calculations are done automatically when the toolpath is imported. The user can also run a visual simulation of the 5-axis 3D printing process by scrolling the simulation slider in our editor to check for potential collisions. Users can change the print head travel height and printing speed at any point during this process. Once the G-code is ready the user can simply export it as a text file and upload it to the machine via the Duet web interface.

\section{Use cases}
\subsection{Support-less 3D printing}
\begin{figure}[h]
  \centering
  \includegraphics[width=1.0\linewidth]{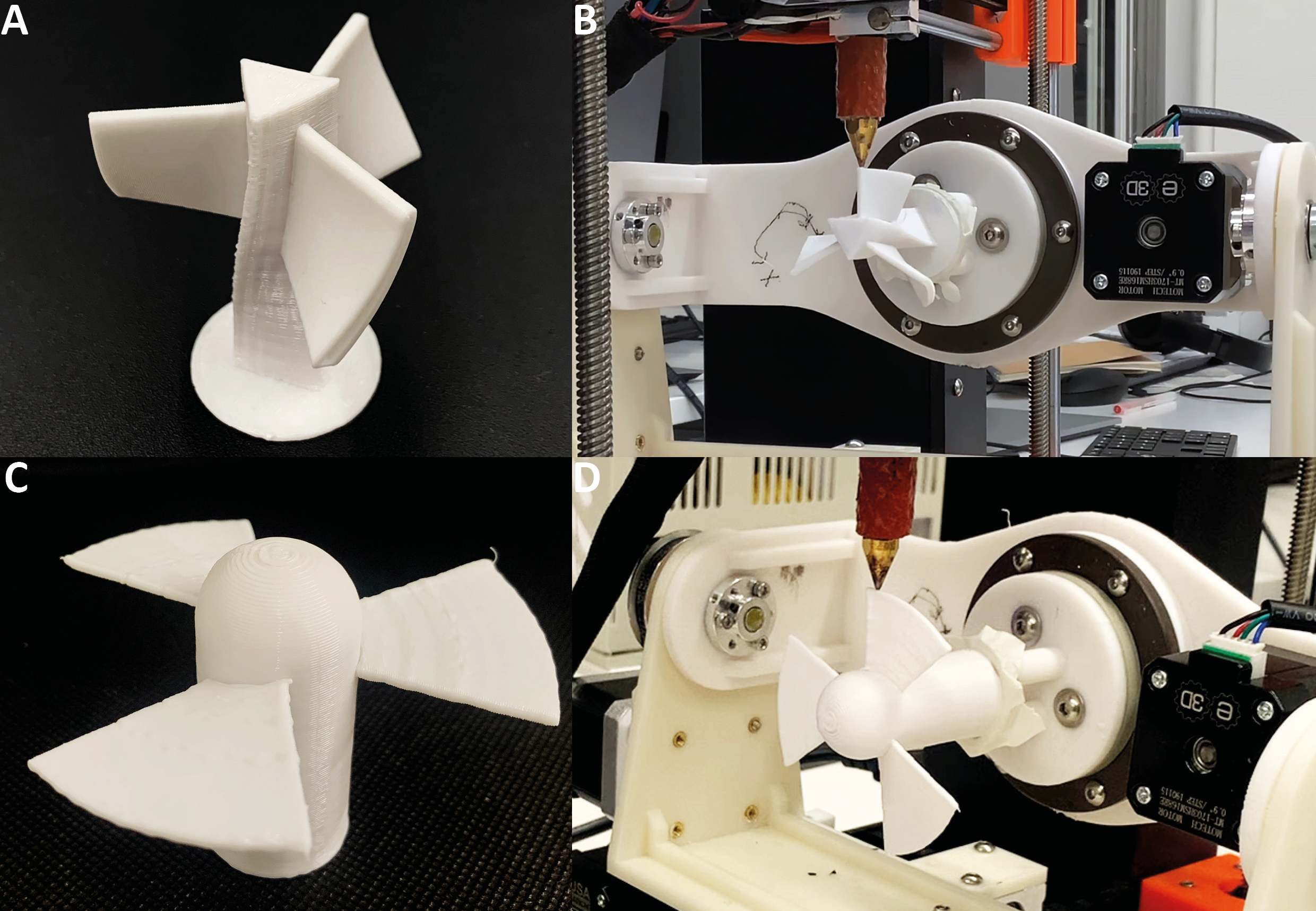}
  \caption{Support-less 3D printing: (a) Printed turbine-like object, and (b) the 3D printing process for that turbine. (c) Printed fan-like object, and (d) the 3D printing process for that fan.}
  \label{overhang}
  \Description{}
\end{figure}

5-axis 3D printing allows support-less extrusion of overhanging geometries. This has several benefits: it reduces the waste of materials caused by extruding the support materials, reduces the print time significantly, removes the oftentimes fiddly and time-consuming need to remove any support material, and results in cleaner surface finishes for the overhanging areas. Figure \ref{overhang} shows a turbine-like and fan-like objects for which the wings were printed conformally, without support.

\subsection{Conformal extrusion}
\begin{figure}[h]
  \centering
  \includegraphics[width=1.0\linewidth]{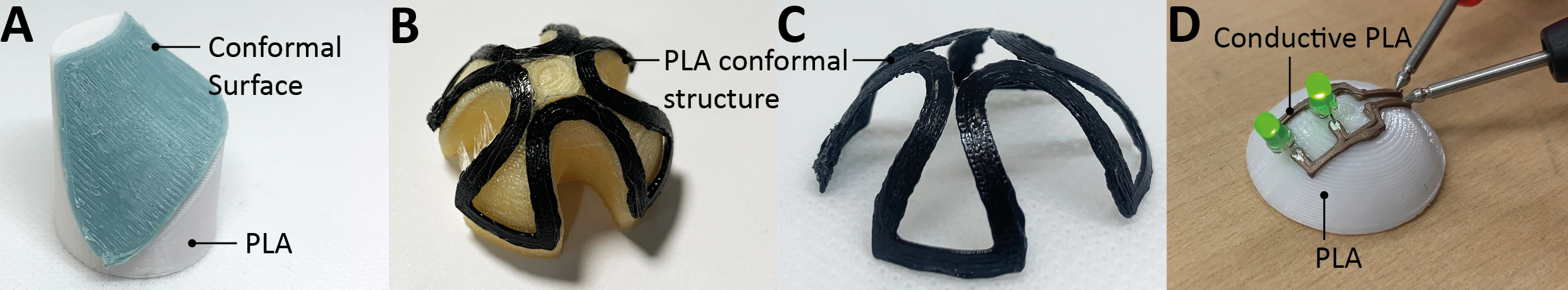}
  \caption{(a) Coloured conformal surface on top of planar 3D printed substrate; (b) conformal structure on top of water-soluble PVA substrate; (c) hollow confomral structure after PVA removal; (d) conductive LED circuit printed conformally onto a soild hemisphere.}
  \label{surfacefinish}
  \Description{}
\end{figure}

Conformal deposition and curved layers allow the user to control the mechanical properties of the 3D printed object by controlling the filament extrusion direction. In the examples shown in Figure \ref{surfacefinish}, we used two different types of filament: one for an underlying `substrate' that was created using regular, 3-axis printing, and another for the conformal layer added via 5-axis printing. In Figure \ref{surfacefinish}(a), the cyan conformal layer is clearly visible. 
Figures \ref{surfacefinish}(b) shows a thin compliant geometry (in black PLA) that was printed on water-soluble material (off-white PVA). In a conventional planar 3D printing setting, constructing such thin free-form inclining structures is challenging due to small overlapping areas between each layers, which leads to weak inter-layer bonding, but a conformal printing approach overcomes this. 

We have also successfully extruded conductive PLA conformally onto a PLA substrate. As discussed by Hong et al. \cite{10.1145/3411764.3445469}, in planar 3D printing conductive PLA has a much higher resistance for an inclined surface compared to purely horizontal due to the additional resistance inherent in inter-layer bonding. Conformal 3D printing overcomes this, because the conductive PLA can be extruded continuously across inclined surfaces at varying angles. This allows greater freedom during the design 3D of printed objects with conductive traces, as illustrated by the hemisphere-shaped LED circuit shown in Figure \ref{surfacefinish}(d). 

\subsection{Structural strengthening}
\begin{figure}[h]
  \centering
  \includegraphics[width=1.0\linewidth]{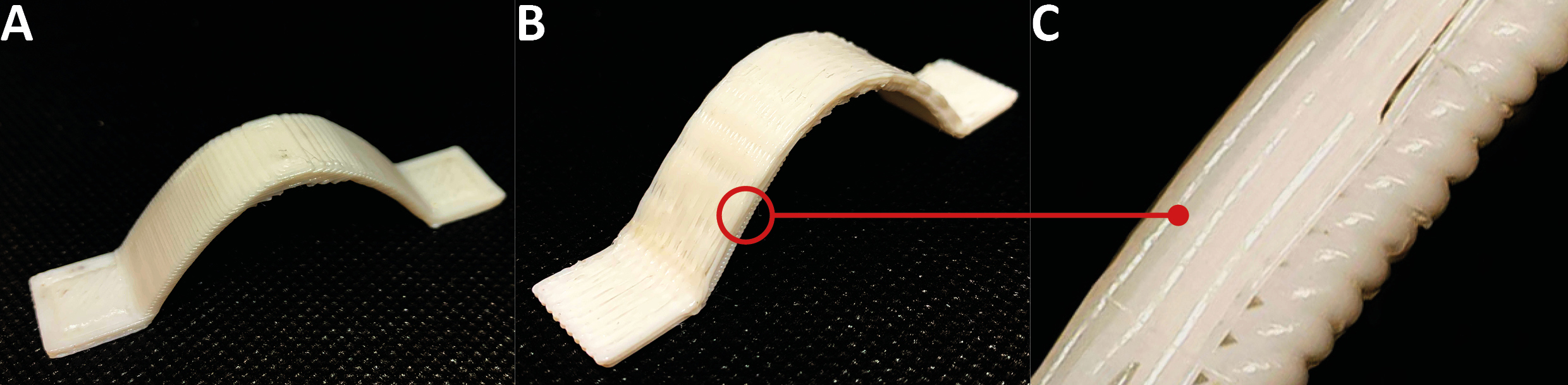}
  \caption{(a) 3D printed bridge by conventional planar construction; (b) Reinforced bridge with mixture of planar 3D printing and conformal 3D printing; (c) Microscopic image of the reinforced layer.}
  \label{bridge}
  \Description{}
\end{figure}

In planar 3D printing, an object is constructed using layer-by-layer deposition that creates anisotropic structure that exhibits significantly weaker inter-layer bondings in comparison to intra-layer bonding. This limits the functional capability of 3D printed mechanical objects due to its weakness or inability to comply with multi-directional loads. Fant et al. and Huang et al. \cite{Fang2020ReinforcedStrength,Huang2015CurvedModelling} have showcased methods of strengthening the mechanical properties of the 3D printed object by reinforcing the object through curved layer depositions. Figure \ref{bridge} shows a simple bridge like structure which was first constructed in planar 3D printing that reinforced with curved layer on top. The overlapping structure between the substrate layer and the conformal layer is shown on Figure \ref{bridge}(c). With conformal reinforcement the user can design cantilevers, hinge and structural parts suitable for applications with multi-directional loads.

\section{Conclusion and future Work}

In this paper we presented a way to convert an existing FFF desktop 3D printer into a 5-axis machine, and demonstrated some of the benefits of 5-axis conformal printing. We have shared detailed step-by-step guidelines for others wishing to adopt our approach in an online repository. For many users, the absence of toolpath generation software is a bigger barrier to 5-axis printing than the machine itself, since it requires an in-depth understanding of kinematics and scripting. Therefore, we have also developed an accessible GUI based conformal slicer that runs within the popular Rhino CAD package. 
This dependency on a commercial product means that our solution is not completely open source, 
but in future work we hope to develop a stand-alone conformal slicer. Going forward we also aim to build a library of 5-axis printable designs for various applications. 

With this work, our primary aim is to lower the barrier of entry to conformal 3D printing. We believe that many hobbyists and makers could benefit from the advantages of low-cost conformal 3D printing showcased in this paper. We also invite other researchers in HCI to explore and build upon the these benefits in future work. In particular, we hope that conformal printing can contribute to research areas such as printed electronics, meta-materials and lightweight lattice structures. We are excited to see what the community will build!


\bibliographystyle{ACM-Reference-Format}
\bibliography{references}

\end{document}